\documentclass[aps,preprint,showpacs,pra]{revtex4}
\usepackage{epsfig,amssymb,amscd}

\begin{document}
\bibliographystyle{prsty}

\title{\bf Selective pulse implementation of two-qubit gates for spin-3/2 based fullerene quantum information processing}
\author{M. Feng$^{1,2,3}$ \footnote[1]{Electronic address: mfeng@thphys.may.ie} and 
J. Twamley$^{1}$ \footnote[2]{Electronic address: jtwamley@thphys.may.ie}} 
\affiliation{$^{1}$ Department of Mathematical Physics, National University of Ireland, Maynooth, Co. Kildare, Ireland \\
$^{2}$ Wuhan Institute of Physics and Mathematics, Chinese Academy of Sciences, Wuhan, 430071, China \\
$^{3}$ School of Theoretical Physics and Department of Applied Physics, Hunan University, Changsha, 410082, China}
\date{\today}

\begin{abstract}

We propose two potentially practical schemes to carry out  two-qubit quantum gates on endohedral fullerenes $N@C_{60}$ or
$P@C_{60}$. The qubits are stored in electronic spin degrees of freedom  of the doped atom $N$ or $P$.
By means of the magnetic dipolar coupling between two neighboring fullerenes, the two-qubit controlled-NOT gate  and the two-qubit 
conditional phase gate are performed by selective microwave pulses assisted by refocusing technique. We will discuss the necessary 
additional steps for the universality of our proposal.
We will also show that our proposal is useful for both quantum gating and the readout of quantum information from the spin-based qubit 
state.
\end{abstract}
\vskip 0.1cm
\maketitle

\section{introduction}

Quantum information processing holds the promise to outperform the computational power of existing computers in the treatment of 
certain problems \cite {cn}. 
Although so far quantum gating has been successfully carried out in some systems, such as trapped ions and nuclear magnetic 
resonance \cite{for}, to have large-scale working quantum information processing a design based on solid-state materials may be 
very promising.

Since both electronic and nuclear spin states typically possess longer decoherence times than charge states, there have been 
a number of proposals for 
solid-state quantum information processing using spin-based qubits, for instance, \cite {kane, rev}.
This work is focused on a spin-based quantum gating with endohedrally doped fullerenes, whose unusual properties have been mentioned
in earlier publications \cite{har,suter,jason}. These fullerenes are highly symmetric hollow molecules which behave as Faraday cages
for the encapsulated atoms $N$ or $P$. The qubits can be encoded in the nuclear spins or the electronic spins of $N$ or $P$, and
the quantum gating in this kind of system is done by using NMR (i.e. Nuclear Magnetic Resonance) or ESR (i.e. Electron Spin Resonance) 
pulse sequences. Since it is less sensitive than 
the electronic spin to decoherence due to environment, the nuclear spin is more suitable for hosting a qubit. However, due to 
larger energy spacing, quantum gating based on electronic spins can be carried out more quickly and readily than that based on 
nuclear spins. Therefore a promising design for quantum information processing has been proposed by encoding qubits into nuclear spins, 
but implementing quantum gates on electronic spins \cite {suter,jason}. The swap operation between 
nuclear and electronic spins via Hyperfine interaction to exchange quantum information is essential to these designs. The 
advantages of this type of quantum information processing device include the lengthy decoherence times of the qubit states due to 
the cage effects of $C_{60}$, and the potentially easier manipulation of the qubits than another proposal \cite{kane}. 

However, to carry out a non-trivial two-qubit gate, say, a controlled-NOT (CNOT) gate, we need at least four steps of manipulation 
with NMR or ESR hard pulse sequences \cite {har,suter,jason}. Although these pulse sequences can be finished within the order of 
$\mu s$, 
since the coupling shifts the energy levels in the two coupled fullerenes with respect to the non-interacting case, the associated 
hard pulses  must  be of wide bandwidths, making their  experimental realization challenging \cite{suter}. To overcome this 
difficulty, we can try to perform gating by selective pulses as discussed in \cite{suter}. 
However, the dopant atom of $N@C_{60}$ or $P@C_{60}$ has three valence electrons, and possesses a quartet ground spin state 
of total spin $3/2$, instead of $1/2$ as simply treated in 
\cite{suter}. As a result, there are a number of degenerate transitions, which makes the problem complicated.
To have a practical gating scheme, we have to first specifically study the true configuration of two coupled $S=3/2$ spins.
Based on the same system as in \cite {har,suter,jason}, we will try to propose two schemes to achieve  non-trivial two-qubit quantum 
gates  with selective pulse operations, which would be useful for both quantum gating and the qubit readout. 

Our proposal is based on electronic spin degrees of freedom of the doped atom N or P, whose total electronic spin is 3/2 with 
four Zeeman levels $|\pm 3/2\rangle$ and $|\pm 1/2\rangle$ in a magnetic field. Throughout the paper, we encode qubits into 
$|\pm 3/2\rangle$, i.e., $|-3/2\rangle \equiv |1\rangle$ and $|3/2\rangle \equiv |0\rangle$. We will show that our schemes are also 
applicable to $|\pm 1/2\rangle$ states. Since the two neighboring fullerenes are 
only distant by the order of {\rm nm}, to achieve a single-qubit operation by individually addressing the qubits, a magnetic 
field gradient has to be introduced. As shown in \cite{suter}, for two nearest-neighbor fullerenes distant by 1.14 {\rm nm} in the 
magnetic field gradient $d B/ dx = 4\times 10^{5}$ {\rm T/m}, the difference between the ESR frequencies of $|\pm 3/2\rangle$ is 
12.7$\times 3 \approx 38$ {\rm MHz}, and thereby the single-qubit operation is possible using modern ESR spectrometers with 
narrow-band pulses. 

When checking the points for universality of our proposal, however, we will have to introduce the nuclear spins of the dopant atoms 
for Hadamard operation. Before going to that point, we will only focus 
on the treatment of electronic spin degrees of freedom.

Our proposal includes two schemes for a CNOT gate and a conditional phase (CPHASE) gate respectively. Both CNOT 
and CPHASE are non-trivial two-qubit gates, which are the essential parts of universal quantum information processing. We will 
first study the configuration of the system of two coupled endohedral fullerenes. Then the main steps of our schemes will be 
presented. We will investigate how to eliminate unwanted relative phases. The usefulness and universality of our proposal will be 
also discussed.   

\section{configuration of two coupled fullerenes}

Consider a system with two $C_{60}$ fullerenes $A$ and $B$, and by neglecting the small terms associated with nuclear spins, we have in 
units of $\hbar=1$,
\begin{equation}
H =  g\mu_{B} B_{A}S_{z}^{A} + g\mu_{B} B_{B} S_{z}^{B} + J S_{z}^{A} S_{z}^{B}
\end{equation}
where $g$ is the electron $g$-factor, $\mu_{B}$ is the Bohr magneton, and $B_{k}$ $(k=A$ and $B)$ is the
magnetic field strength sensed by fullerene $k$. $J$ is the magnetic dipolar coupling strength between  
neighboring fullerenes. As in \cite{suter}, we suppose  $J\sim 50$ {\rm MHz}.
$$S_{z}^{A}=\frac {1}{2}\pmatrix {3 & 0 & 0 & 0 \cr 0 & 1 & 0 & 0 \cr 0 & 0 & -1 & 0 \cr  0 & 0 & 0 & -3} 
\otimes I_{B}$$ and $$S_{z}^{B}= I_{A}\otimes  \frac {1}{2} \pmatrix
{3 & 0 & 0 & 0 \cr 0 & 1 & 0 & 0 \cr 0 & 0 & -1 & 0 \cr  0 & 0 & 0 & -3}.$$
Direct calculation shows that in the space spanned by  
\begin{quote}
\begin{tabular}{rrrr}
$|\frac {3}{2}, \frac {3}{2} \rangle,$ & $|\frac {3}{2}, \frac {1}{2} \rangle,$ &
$|\frac {3}{2}, -\frac {1}{2} \rangle,$ & $|\frac {3}{2}, -\frac {3}{2} \rangle,$ \\ 
$|\frac {1}{2}, \frac {3}{2} \rangle,$ & $|\frac {1}{2}, \frac {1}{2} \rangle,$ &
$|\frac {1}{2}, -\frac {1}{2} \rangle,$ & $|\frac {1}{2}, -\frac {3}{2} \rangle,$ \\
$|-\frac {1}{2}, \frac {3}{2} \rangle,$ & $|-\frac {1}{2}, \frac {1}{2} \rangle,$ &
$|-\frac {1}{2}, -\frac {1}{2} \rangle,$ & $|-\frac {1}{2}, -\frac {3}{2} \rangle,$ \\
$|-\frac {3}{2}, \frac {3}{2} \rangle,$ & $|-\frac {3}{2}, \frac {1}{2} \rangle,$ &
$|-\frac {3}{2}, -\frac {1}{2} \rangle,$ & $|-\frac {3}{2}, -\frac {3}{2} \rangle,$ 
\end{tabular}
\end{quote}
the eigenenergies are respectively,
\begin{quote}
\begin{tabular}{rrrr}
$3\omega_{1}+ 3\omega_{2}+ 9J/4,$ & $3\omega_{1}+ \omega_{2}+ 3J/4,$ & $3\omega_{1} -\omega_{2}-  3J/4,$ & 
$3\omega_{1}- 3\omega_{2}- 9J/4,$ \\ 
$\omega_{1}+ 3\omega_{2}+ 3J/4,$ & $\omega_{1}+ \omega_{2}+ J/4,$ & $\omega_{1}- \omega_{2} -J/4,$ & $\omega_{1}- 3\omega_{2} -3J/4,$\\ 
$-\omega_{1}+ 3\omega_{2}-3J/4,$ & $-\omega_{1}+ \omega_{2} -J/4,$ & $-\omega_{1}-\omega_{2}+ J/4,$ & $-\omega_{1}- 3\omega_{2}+ 3J/4,$\\
$-3\omega_{1}+ 3\omega_{2}- 9J/4,$ & $-3\omega_{1}+ \omega_{2}-3J/4,$ & $-3\omega_{1}-\omega_{2}+ 3J/4,$ & 
$-3\omega_{1}-3\omega_{2}+ 9J/4,$
\end{tabular}
\end{quote}
where $\omega_{1} = g\mu_{B}B_{A}/2$, and $\omega_{2} = g\mu_{B}B_{B}/2$. It is obvious that each level has shifted from it's
original position due to the magnetic dipolar coupling, and there are numerous degenerate transitions
in this two spin-3/2 system, as shown in Fig. 1. This is an important difference between systems comprising of two spin-1/2 particles 
and two spin-3/2 particles.  

\section{two-qubit quantum gating}

\subsection{CNOT Gate}

CNOT is the most commonly used two-qubit quantum gate. Our implementation is based on a characteristic of the 
configuration shown in Fig. 1, that is, the degenerate transition frequency of a spin state is heavily dependent on the 
coupled (or neighboring) spin state. Due to degeneracy and the magnetic field gradient, the radiation of a ESR pulse on a single qubit 
yields following Hamiltonian in units of 
$\hbar=1$,
\begin{equation}
H = \omega_{0} S_{z} + \Omega \left( e^{-i\omega_{L}t} S_{+} + e^{i\omega_{L}t} S_{-}\right)
\end{equation}  
where $\omega_{0}$, different from the non-interacting counterpart, is one of the degenerate transition frequencies labeled in Fig. 1.
$\omega_{L}$ is the frequency of the microwave pulse.  $S_{k}$ $(k=z, +,$ or $-)$ is a 4$\times$4 Pauli operator. $\Omega$ is the
Rabi frequency. When $\omega_{L}=\omega_{0}$, in the interaction Hamiltonian we have $H_{I} = \Omega S_{x}$ with 
$$S_{x}=\frac {1}{2}\pmatrix {0 & \sqrt{3} & 0 & 0 \cr \sqrt{3} & 0 & 2 & 0 \cr 0 & 2 & 0 & \sqrt{3} \cr  0 & 0 & \sqrt{3} & 0}.$$
For a $\pi$ pulse radiation of ESR, i.e. $\Omega t=\pi$, $H_{I}$ yields operator
$$\hat{P}=i\pmatrix {0 & 0 & 0 & 1 \cr 0 & 0 & 1 & 0 \cr 0 & 1 & 0 & 0 \cr  1 & 0 & 0 & 0},$$ which works independently in the 
subspace spanned by $|\pm 3/2\rangle$ or the one spanned by $|\pm 1/2\rangle$.

Therefore, with the operator $\hat{P}$, we can flip states $|\pm 3/2\rangle$ of a single qubit with a ESR pulse whose frequency is 
determined by the neighboring spin state. This is actually a CNOT operation. Since with the magnetic field gradient we are able to 
flip different spin states with different ESR frequencies, we can perform different CNOT gates. For example, with the frequency 
$2\omega + \Delta -3J/2$, we have a CNOT$_{AB}$, i.e., the target spin state 
in fullerene B flipped only in the case when the control qubit is $|-3/2\rangle_{A}$. While to have a CNOT$_{BA}$, we use the ESR 
pulse with the frequency $2\omega -3J/2$. 

Given a perfect experimental implementation, the fidelity of our scheme only depends on the exact knowledge of the configuration
of the considering system. The implementation time is determined by the Rabi frequency $\Omega$.  

\subsection{CPHASE Gate}

CPHASE gate is another useful two-qubit gate, which plays an important role in Grover search to flip the phase of the 
labeled state \cite{grover}. CPHASE gate is always considered to be equivalent to CNOT if single qubit rotation is easily performed. 
But in the system under consideration, single qubit gating is not an easy  job (see Sec IV). So it is interesting to have a 
straightforwardly produced CPHASE gate.
The key idea of our scheme is to radiate the two $C_{60}$ fullerenes by using two detuned microwaves, which is similar to what is done
in ion trap quantum computing proposals \cite {sm}. However, the fullerene problem under consideration is different from atomic 
problems 
in \cite {sm}. First, there is no vibrational degrees of freedom attached to the qubit states. Secondly, we are considering a 
spin-3/2 system, which is more complicated. Thirdly, the two fullerenes under the magnetic field gradient possess different  
transition resonance frequencies for the two qubit states. 

We employ spin states $|-1/2\rangle_{A(B)}$ as auxiliary states, and couple the states $|-3/2\rangle$ and $|-1/2\rangle$ of the two 
endohedrals by two detuned microwaves. Keep in mind that although the computational subspace is spanned by $|\frac {3}{2}\rangle_{A(B)}$
and $| \frac {-3}{2}\rangle_{A(B)}$, the subspace under 
consideration is spanned by $|\frac {-1}{2}\rangle_{A(B)}$, $|\frac {3}{2}\rangle_{A(B)}$ and $| \frac {-3}{2}\rangle_{A(B)}$.
From the above eigenenergies, we know that $\omega_{-1/2,-1/2} - \omega_{-3/2,-3/2}= 
2 \omega_{1} + 2 \omega_{2} - 2 J$, $\omega_{A}=\omega_{-1/2,-3/2} - \omega_{-3/2,-3/2} = 2\omega_{1} -3J/2$ and 
$\omega_{B}=\omega_{-3/2,-1/2} - \omega_{-3/2,-3/2} = 2\omega_{2} - 3J/2$. So as shown in Fig. 2, an additional shift of 
J is suffered by $|\frac {-1}{2}, \frac {-1}{2}\rangle$. Since the inter-fullerene spacing is only 1.14 {\rm nm}, we consider
two in-plane directed  microwaves, which can be described as 
$\cos (\omega_{sl}t + \phi_{l})\hat{e}_{x} + \sin (\omega_{sl}t + \phi_{l})\hat{e}_{y}$ with $\omega_{sl}$ the frequency of the 
microwave pulse, $\phi_{l}$ the phase of the microwave and $l=1, 2$. These radiate the two fullerenes simultaneously. The 
Hamiltonian for such a system can be written in units of $\hbar=1$ as $H=H_{0} + H_{I}$ where
\begin{equation}
 H_{0}= J |\frac {-1}{2}, \frac {-1}{2}\rangle\langle \frac {-1}{2}, \frac {-1}{2}| +
 \frac {\omega_{A}}{2} \left(|\frac {-1}{2}\rangle\langle \frac {-1}{2}| - 
|\frac {-3}{2}\rangle\langle \frac {-3}{2}|\right)_{A}\otimes I_{B} +
\frac {\omega_{B}}{2} I_{A}\otimes \left(|\frac {-1}{2}\rangle\langle \frac {-1}{2}|
- |\frac {-3}{2}\rangle\langle \frac {-3}{2}|\right)_{B} 
\end{equation} 
\begin{equation}
H_{I}= \frac {\Omega_{A}}{2} \left[e^{i(\omega_{s1}t+\phi_{1})} + e^{i(\omega_{s2}t+\phi_{2})}\right]|\frac {-3}{2}\rangle_{A}
\langle\frac {-1}{2}|
\otimes I_{B} + \frac {\Omega_{B}}{2} \left[e^{i(\omega_{s1}t+\phi_{1})} + e^{i(\omega_{s2}t+\phi_{2})}\right]I_{A}\otimes 
|\frac {-3}{2}\rangle_{B}\langle \frac {-1}{2}| + h.c.
\end{equation} 
where $I_{k}= (|\frac {-1}{2}\rangle\langle \frac {-1}{2}| + |\frac {-3}{2}\rangle\langle \frac {-3}{2}|)_{k}$. For simplicity, 
we have supposed that $\phi_{l}$ is the same for two fullerenes 
experiencing the same microwave, and the coupling strength $\Omega_{k}$ is identical for each fullerene irradiated by the separate
microwave sources. In the rotating frame with respect to $H_{0}$, we have 
$$ H_R = \frac {\Omega_{A}}{2} \left[ e^{i(\omega_{s1}t+\phi_{1})} + e^{i(\omega_{s2}t+\phi_{2})}\right] e^{-i\omega_{A} t} 
|\frac {-3}{2}\rangle_{A}\langle\frac {-1}{2}|\otimes \left(|\frac {-1}{2}\rangle\langle \frac {-1}{2}| e^{-iJ t} + 
|\frac {-3}{2}\rangle\langle \frac {-3}{2}|\right)_{B} + $$
\begin{equation}
\frac {\Omega_{B}}{2} \left[ e^{i(\omega_{s1}t+\phi_{1})} + e^{i(\omega_{s2}t+\phi_{2})}\right] e^{-i\omega_{B} t} 
\left(|\frac {-1}{2}\rangle\langle \frac {-1}{2}| e^{-iJ t} + |\frac {-3}{2}\rangle\langle \frac {-3}{2}|\right)_{A} 
\otimes |\frac {-3}{2}\rangle_{B}\langle\frac {-1}{2}|+ h.c.
\end{equation}
Since the two qubits are radiated simultaneously by two detuned microwave pulses,
there should be four different detunings, say, $\delta_{1}$, $\delta_{2}$, $\delta_{3}$ and $\delta_{4}$, with 
$\delta_{1}+\delta_{3}= \delta_{2}+\delta_{4} = J$ to achieve resonance. So we consider this scheme to be also a selective pulse method.
If we can make {\rm max} $\{\Omega_{A}, \Omega_{B}\} \ll 2$ {\rm min} $\{|\delta_{1}|, |\delta_{2}|, |\delta_{3}|, |\delta_{4}| \}$, 
then we will force effective transitions between $|\frac {-1}{2}, \frac {-1}{2}\rangle$, and $|\frac {-3}{2}, \frac {-3}{2}\rangle$,
via two virtually occupied intermediate states $|\frac {-1}{2}, \frac {-3}{2}\rangle$ and $|\frac {-3}{2}, \frac {-1}{2}\rangle$ 
by second-order perturbative expansion \cite{sm,ion}. The effective Hamiltonian is thus
\begin{equation}
\tilde{H} = \frac {\tilde{\Omega}}{2} \left(|\frac {-3}{2}, \frac {-3}{2}\rangle\langle \frac {-1}{2}, 
\frac {-1}{2}| e^{i(\phi_{1}+\phi_{2})}+ h.c.\right) 
\end{equation}
where
\begin{equation}
\frac {\tilde{\Omega}}{2} = \frac {1}{4} \Omega_{A}\Omega_{B} \left(\frac {1}{\delta_1} + \frac {1}{\delta_2} +
\frac {1}{\delta_3} +\frac {1}{\delta_4}\right),
\end{equation}
$\delta_{1}= \omega_{s1}-\omega_{A}$, $\delta_{2}= \omega_{s1}-\omega_{B}$, $\delta_{3}= \omega_{s2}-\omega_{B}$, 
and $\delta_{4}= \omega_{s2}-\omega_{A}$. Returning to the Schr\"odinger representation, the time evolutions based on Eq. (6) are 
\begin{equation}
|\frac {-3}{2}, \frac {-3}{2}\rangle \rightarrow \cos (\tilde{\Omega}t/2)|\frac {-3}{2}, \frac {-3}{2}\rangle 
- i e^{-i(\omega_{A}+\omega_{B}+J)t}e^{-i(\phi_{1}+\phi_{2})}\sin (\tilde{\Omega}t/2)|\frac {-1}{2}, \frac {-1}{2}\rangle ,
\end{equation}
\begin{equation}
|\frac {-1}{2}, \frac {-1}{2}\rangle \rightarrow  e^{-i(\omega_{A}+\omega_{B}+J)t}\cos (\tilde{\Omega}t/2)
|\frac {-1}{2}, \frac {-1}{2}\rangle - i e^{i(\phi_{1}+\phi_{2})}\sin (\tilde{\Omega}t/2)|\frac {-3}{2}, \frac {-3}{2}\rangle .
\end{equation}
Our two-qubit CPHASE gate is carried out by Eq. (8). Since $|\frac {-1}{2}, \frac {-1}{2}\rangle$ is out of our computing subspace, 
and is not populated initially, if we implement a $2\pi$-pulse of microwave radiation, i.e. $\tilde{\Omega} t = 2\pi$,
we will have  $|\frac {-3}{2}, \frac {-3}{2}\rangle \rightarrow -|\frac {-3}{2}, \frac {-3}{2}\rangle$, but no change for other qubit 
states in the computational subspace. This is a typical  CPHASE gate with the form 
$|\alpha, \beta\rangle\rightarrow e^{i\alpha\beta\pi}|\alpha, \beta\rangle$, where $\alpha$ and $\beta$ are the logic state 0 or 1
respectively.  To achieve this CPHASE gate, we should make sure that the implementation time is shorter than the 
decoherence time of the electronic spin. Since $\delta_{1}+\delta_{3} = \delta_{2}+\delta_{4}= 50$ {\rm MHz} and 
$\omega_{B}-\omega_{A}=12.7$ {\rm MHz}, Eq. (6) can be rewritten as 
\begin{equation}
\frac {\tilde{\Omega}}{2} = \frac {1}{4} \Omega_{A}\Omega_{B} \left(\frac {1}{\delta_1} + \frac {1}{\delta_1 - 12.7} +
\frac {1}{50- \delta_1} +\frac {1}{62.7-\delta_1}\right).
\end{equation}
To have a two-qubit CPHASE gate with high fidelity, we introduce $P=\Omega_{0}/2\delta_{min}$ with 
$\Omega_{0}= ${\rm max} $\{\Omega_{A}, \Omega_{B}\}$ and $\delta_{min}=$ {\rm min}  $\{|\delta_1|, |\delta_1 - 12.7|, |50- \delta_1|, 
|62.7-\delta_1| \}$.  The excitation probability of the intermediate states is defined as $\tilde{n}=2P^{2}$ \cite {ion}. 
The smaller the value of $\tilde{n}$, the higher the fidelity of our gate, but the longer the implementation time, as shown in 
Fig. 3. From Fig. 3, we also know that the shortest implementation time occurs in detuning $\delta_{1}=31.35$ {\rm MHz}.

\section{discussion}

With current experimental techniques, Rabi frequencies of 20 $\sim$ 30 MHz are available 
\cite {meyer}, which fully meets the requirements of our schemes. However, compared to our CNOT gate scheme,  our CPHASE gate works 
somewhat slowly due to the 
detuning and weak interaction. It takes time of the order of $\mu s$, which is only slightly 
shorter than $T_{2}$ $(\sim 20 \mu s)$ of the electronic spin at low temperature ($\sim$ $7^{o}$ Kelvin, high spin concentration) 
\cite {kno}. So, in order to carry out this scheme, we need an increased $T_{2}$. With decoupling, and decreasing spin density, 
$T_{2}$ will be increased to $T_{1}$ $(\sim 1~ sec)$ \cite {jason}. The expectation of $T_{2}$ prolonged 
toward $T_{1}$ is essentially not only to our proposal, but also to all the other spin-based quantum information processing schemes.  
If $T_{2}$ can be as long as 1 $sec$, millions of our CPHASE gates will be available within the decoherence time of the electronic spin.

The numerical simulation for the CPHASE gate also shows that the smaller the difference between $\omega_{1}$ and $\omega_{2}$, the 
shorter the 
implementation time of our gating. The shortest implementation time is 1.54 $\mu s$ in the case of $\omega_{1}=\omega_{2}$ and 
$\delta_{1}= \delta_{2}= \delta_{3}= \delta_{4}= J/2$, which corresponds to the situation of $dB / dx = 0$. This can be understood in  
that the optimal operation of our gate corresponds to the same detunings of the two microwaves for two exactly identical fullerenes.   
However, in the architecture proposed in \cite {har,suter,jason}, a magnetic field gradient is necessary for single qubit operation. 
What is the optimal field gradient depends on the fluctuations of the magnetic 
field, our capability to distinguish different fullerenes, and $T_{2}$ of the electronic spins.

\subsection{Single Qubit Operation}

So far, we have only discussed the non-trivial portion of the two-qubit operation. Before discussing the rephasing of the trivial system
dynamics, it is essential to note that to carry out universal quantum information processing we need single qubit 
rotation besides the above mentioned two-qubit gates. Due to the magnetic dipolar coupling, as mentioned above, the transition 
frequencies between the two qubit states are dependent on the relative spin states. This enables CNOT gates but complicates the 
implementation of single qubit rotations. Fortunately, the operator $\hat{P}$ can be carried out independently in $|\pm 3/2\rangle$ 
or $|\pm 1/2\rangle$. So in the presence of the magnetic field gradient we can perform the single qubit operation by simultaneously 
irradiating with two selective ESR pulses. For example, to flip a single spin state in fullerene B, we use ESR pulses with frequencies 
$2\omega + \Delta -3J/2$ and 
$2\omega + \Delta +3J/2$. If the Rabi frequency is 25 MHz, the implementation time would be 0.126 $\mu$s.

However, we have no way to execute a Hadamard gate on the $|\pm 3/2 \rangle$ subspace within our model. To this end, we have to 
introduce nuclear spins, as done in \cite{suter,jason}. In most doped fullerene proposals for quantum information processing the qubits 
are encoded in nuclear spins and the electronic spins are only employed for quantum gating, as mentioned in Sec I above. Therefore, 
with fullerenes containing both nuclear and electronic spins, our schemes incorporating  
selective pulse operations for two-qubit gates on electronic spins will suffice to yield universal quantum information processing.

\subsection{Removal of relative phases}

What we have proposed in section III above however, is the nontrivial portion of the specified two-qubit gates with one-step 
manipulation. If we consider a general case, i.e. the initial 
two-qubit state to be $a|3/2, 3/2\rangle+ b |3/2, -3/2\rangle + c |-3/2, 3/2\rangle +d |-3/2, -3/2\rangle$, our schemes could not be 
carried out so simply because additional relative phases will appear in the superposition during the gating due to free evolution. 
To get rid of these trivial phases, we have to use a refocusing pulse sequence of $\pi/2$ 
hard ESR pulses \cite {cn}. 

For instance, within $\tau$, we have finished a CPHASE gate with our scheme. In contrast to 
the desired state $\Psi_{1}= a|3/2, 3/2\rangle+ b |3/2, -3/2\rangle + c |-3/2, 3/2\rangle - d |-3/2, -3/2\rangle$, 
we actually have $\Psi_{2}= a e^{-i\theta_{1}}|3/2, 3/2\rangle+ b e^{-i\theta_{2}}|3/2, -3/2\rangle + c e^{-i\theta_{3}}|-3/2, 
3/2\rangle - d |-3/2, -3/2\rangle$, where $\theta_{1}=(12\omega + 3\Delta)\tau$, $\theta_{2}=(6\omega + 3\Delta - 9J/2)\tau$, 
$\theta_{3}=(6\omega -9J/2)\tau$, and we omitted the global phase. By making use of 
$e^{i\pi S_{x}} S_{z} e^{-i\pi S_{x}}=- S_{z}$, we can remove the relative phases by
sequentially sending $\pi/2$ hard ESR pulses to specific electronic spins of the fullerenes and then waiting for $\tau$. As shown 
in Table I, the undesired phases are eliminated by continuously carrying out the three steps. The removal of relative phases for 
our CNOT gating can be done analogously.
 
Although our scheme is more complicated with the inclusion of this rephasing pulse sequence, our schemes could perform quantum 
computing correctly, and the implementation is not more difficult than  in \cite {suter,jason}. Moreover, the refocusing is a mature
technique, which has been widely applied in various experiments of NMR \cite {cn}. Furthermore, we will show below that
our schemes would be also very useful for the last stage of quantum computing, i.e. the qubit readout.  

\subsection{Readout}

The readout problem in spin-based solid state quantum computing is still an open question \cite {jason1}. For fullerene - based 
quantum information processing, possible methods to read out the states of the qubit may be to use Magnetic Resonance Force Microscopy 
\cite {berman} or spin-dependent Single Molecule Transistor \cite {park}. Achieving single spin sensitivities in such methods will 
require significant developments.
Another promising technology for spin state detection is Micro-SQUID (superconducting quantum interference device). 
It had been reported that this device is able to detect the spin state in systems with $S \ge 10$ \cite {pakes,wern}. So by coupling
a nanomolecular magnet $Fe_{8}$ or $Mn_{12}$ with total spin $S=10$ \cite{wern} to an adjacent fullerene in a magnetic field, 
we can attempt a SWAP or CNOT 
between $|\pm 3/2\rangle$ states of our fullerene and $|\pm 10\rangle$ states of the nanomolecular magnet by our selective pulse scheme,
along with quantum tunneling of magnetization \cite {ft}. Since the spin states 
should be well polarized in the readout stage of quantum information processing, no relative phase would appear 
during the gating and thereby no repair step should be taken. So the CNOTs could be carried out by one-step operations.
After the information is converted from $|\pm 3/2\rangle$ to $|\pm 10\rangle$ \cite{exp},  our readout 
can be achieved by detecting $|\pm 10\rangle$ with an array of Micro-SQUIDs.

\section{conclusion}

To achieve our schemes, we need stable and homogeneous magnetic field gradients as well as reliable 
radiation sources. As stated in \cite {suter}, the desired magnetic field gradient can be provided by currents through 
nanometer-sized wires. Experimentally, high quality ESR pulses have been widely applied, and
utilizing two detuned radiation sources is already a sophisticated technique \cite {ion}. With current $T_{2}$ of the electronic spins, 
our proposed CNOT can be carried out with fidelity of $100\%$ by more than 100 times, and
our CPHASE with the fidelity of $95\%$ can be achieved coherently approximately eight times.

Although our discussion has been focused on qubit states $|\pm 3/2\rangle$, our proposal is easily applicable to 
$|\pm 1/2\rangle$ encoding qubits. In this case, for example,  the CPHASE gate works with $| 3/2\rangle$ to be the auxiliary state.
Moreover, as discussed in \cite{jason}, with a nonlinear term in the spin Hamiltonian, for example, a Zero-Field-Splitting term 
$\sim S_{z}^{2}$, one can  lift the degeneracy and can execute a Hadamard gate on qubit states $|\pm 1/2\rangle$ using selective 
excitations. Furthermore the rephasing  steps  are identical to the above case of $|\pm 3/2\rangle$. So universal quantum 
information processing is available in principle in this case too.  

In conclusion, two potentially practical two - qubit gates  have been proposed for spin-based quantum information processing with 
endohedral fullerenes. Since we considered a true system with coupling spin -3/2 components, the situation is more complicated 
than in \cite{suter}. By using selective pulses and excluding hard, wideband pulses, we can reduce the susceptibility to decoherence.
We have discussed how to efficiently implement the selective pulse schemes and shown that our schemes are alternative ways
to two-qubit gating in fullerene based systems. Furthermore, it can be easily 
checked that our proposal, based on the pairwise coupling, i.e. the magnetic dipolar coupling, can be generalized
to the multi-fullerene case, where our proposed quantum gates can be implemented in parallel.  
Based on the above discussion, we argue that
our proposal is not only useful for quantum gating, but also advantageous in the achievement of single spin detection.

\section{acknowledgments}

The authors thank Jim Greer, Wolfgang Harneit, Carola Meyer, Wolfgang Wernsdorfer and the referee for helpful discussion. MF is 
grateful for support 
from Chinese Academy of Sciences. The work is supported by EU Research Project QIPDDF-ROSES under contract number IST-2001-37150.


\newpage

\vspace{20 pt}

{\bf Table I}. The refocusing steps for removing relative phases, where $\theta_{1}$, $\theta_{2}$ and $\theta_{3}$ are 
the undesired phases related to $|3/2,3/2\rangle$, $|3/2,-3/2\rangle$ and $|-3/2,3/2\rangle$, respectively. $[S_{k}]$ means 
$\exp {(-i\pi S_{x}^{k})}$ with $k= A, B$. H is given in Eq. (1).

\vspace{20 pt}

\begin{tabular}{|l|l|l|}
\hline
Step & ~~~~~~~~~~ Refocusing ~~~& Relative phases \\
\hline
0 &  & $\theta_{1}$ ~~~ $\theta_{2}$ ~~~ $\theta_{3}$ ~~~ \\
\hline
1 & $W_{1}=[-S_{x}^{A}]\exp {(-iH\tau)}[S_{x}^{A}]$ & 0~~~~~ 0~~~~ $2\theta_{3}$ \\
\hline
2 & $W_{2}=[-S_{x}^{B}]\exp {(-iH\tau)}[S_{x}^{B}]$ & $-\theta_{1}$~~ $\theta_{2}$~~~ $\theta_{3}$ \\
\hline
3 & $W_{3}=[-S_{x}^{A}][-S_{x}^{B}]\exp {(-iH\tau)}[S_{x}^{B}][S_{x}^{A}]$ & 0~~~~~ 0~~~~~ 0~~~ \\
\hline
\end{tabular}

\newpage

\begin{figure}
\begin{center}
\setlength{\unitlength}{1cm}
\begin{picture}(6,8)
\put(-3.5,0){\includegraphics[width=10cm]{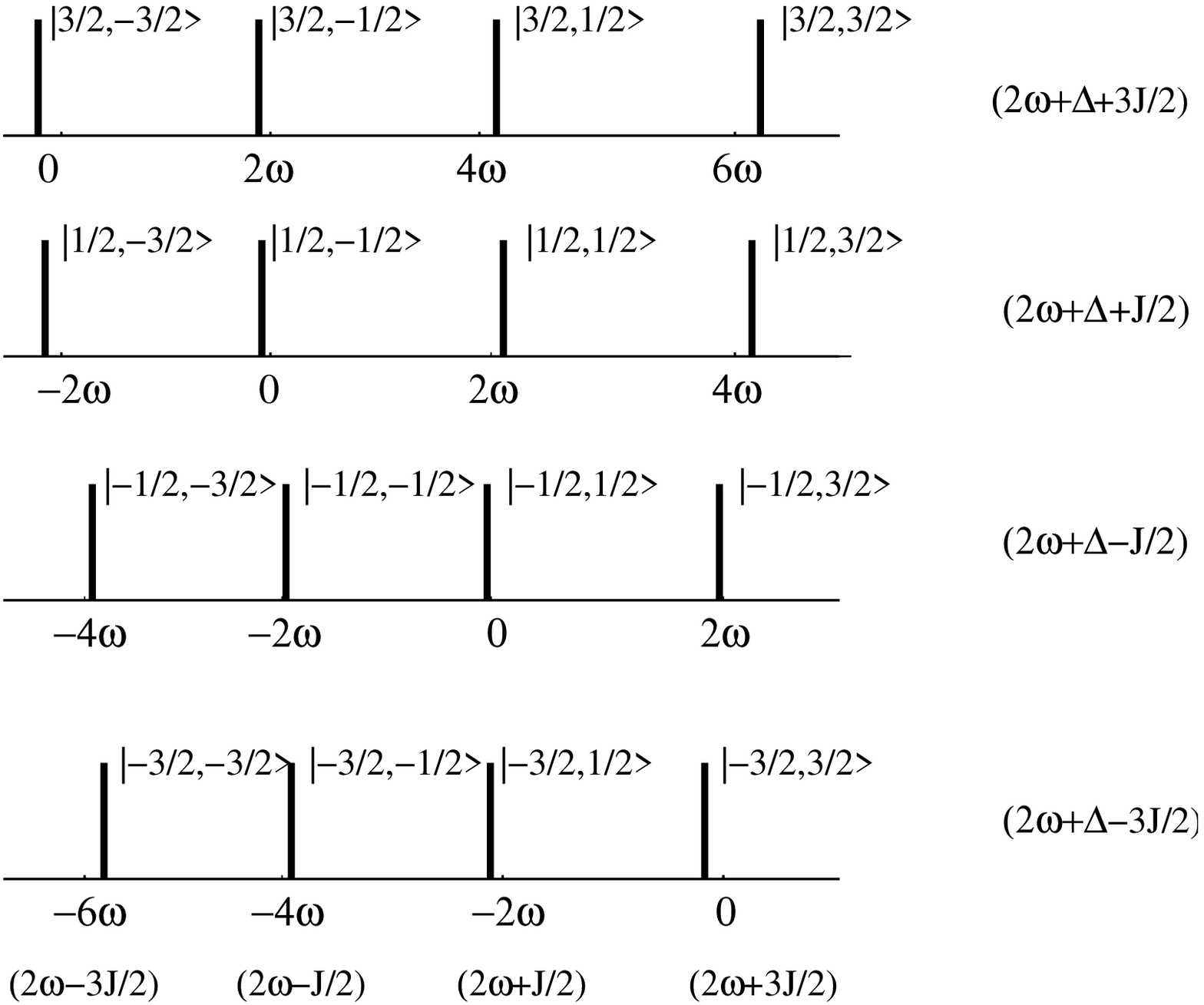}}
\put(0,9){\Large (a)}
\end{picture}
\end{center}
\label{Fig1a}
\end{figure}

\begin{figure}[p]
\begin{center}
\setlength{\unitlength}{1cm}
\begin{picture}(6,10)
\put(-3.5,0){\includegraphics[width=11cm]{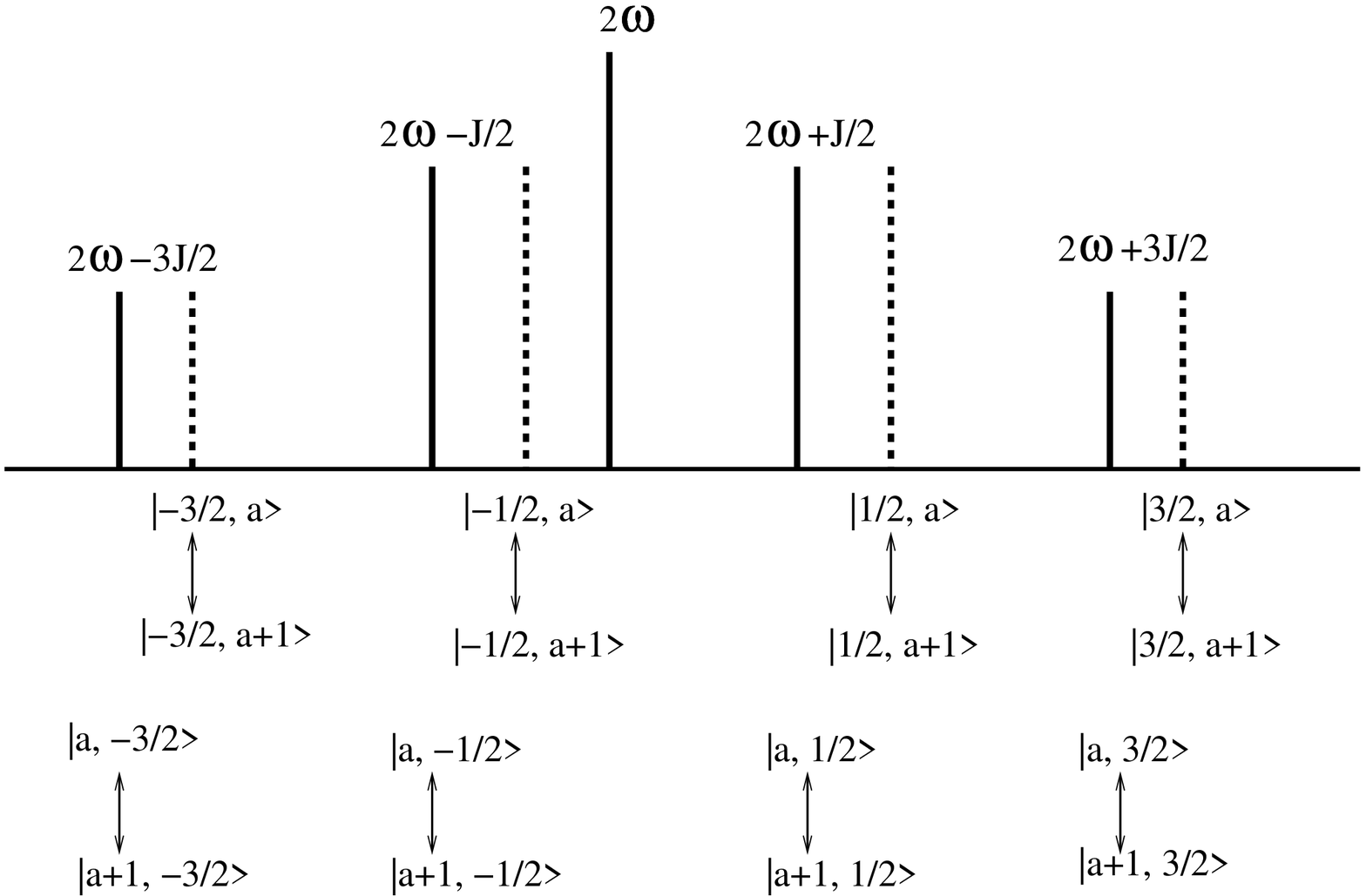}}
\put(0,8){\Large (b)}
\end{picture}
\end{center}
\caption{ (a) The spectrum of the eigenenergies of the two fullerene system coupled by the magnetic dipole-dipole interaction,
where $\omega=\omega_{1}$ and $\Delta=2\omega_{2}-2\omega_{1}=12.7$ $MHz$. $(\cdots)$ represents
the degenerate frequency difference between the nearest-neighbor levels in a row or 
in a column. (b) The spectrum of transition frequency corresponding to (a), where $a=-3/2, -1/2$ and 1/2. The dashed lines
correspond to the shift of the solid counterparts by $\Delta$.}
\label{Fig1b}
\end{figure}

\begin{figure}[p]
\begin{center}
\setlength{\unitlength}{1cm}
\begin{picture}(6,8)
\put(-3.5,0){\includegraphics[width=10cm]{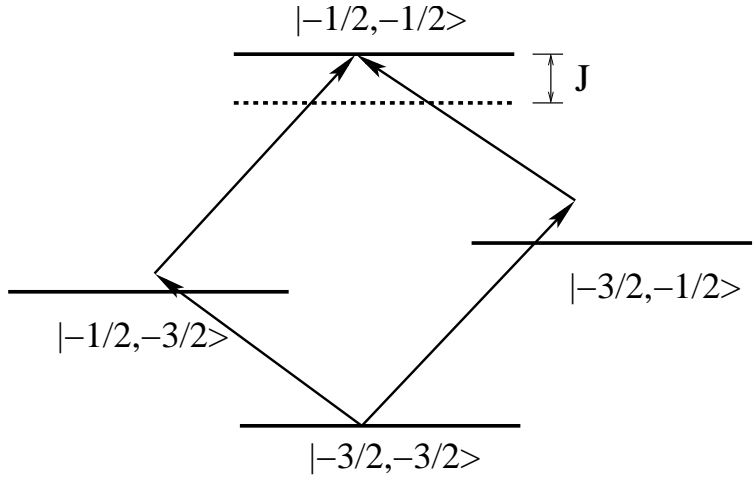}}
\end{picture}
\end{center}
\caption{ Two neighboring fullerenes $A$ and $B$, radiated simultaneously by two in-plane directed microwaves, coupling
spin states $|-1/2\rangle$ and $|-3/2\rangle$.  The two-photon process is for resonant transition between
$|-1/2, -1/2\rangle$ and $|-3/2, -3/2\rangle$, where $|-1/2, -3/2\rangle$ and $|-3/2, -1/2\rangle$ are non-populated intermediate states
due to large detuning and weak radiation coupling.}
\label{Fig2}
\end{figure}

\begin{figure}[p]
\begin{center}
\setlength{\unitlength}{1cm}
\begin{picture}(6,10)
\put(-3.5,0){\includegraphics[width=10cm]{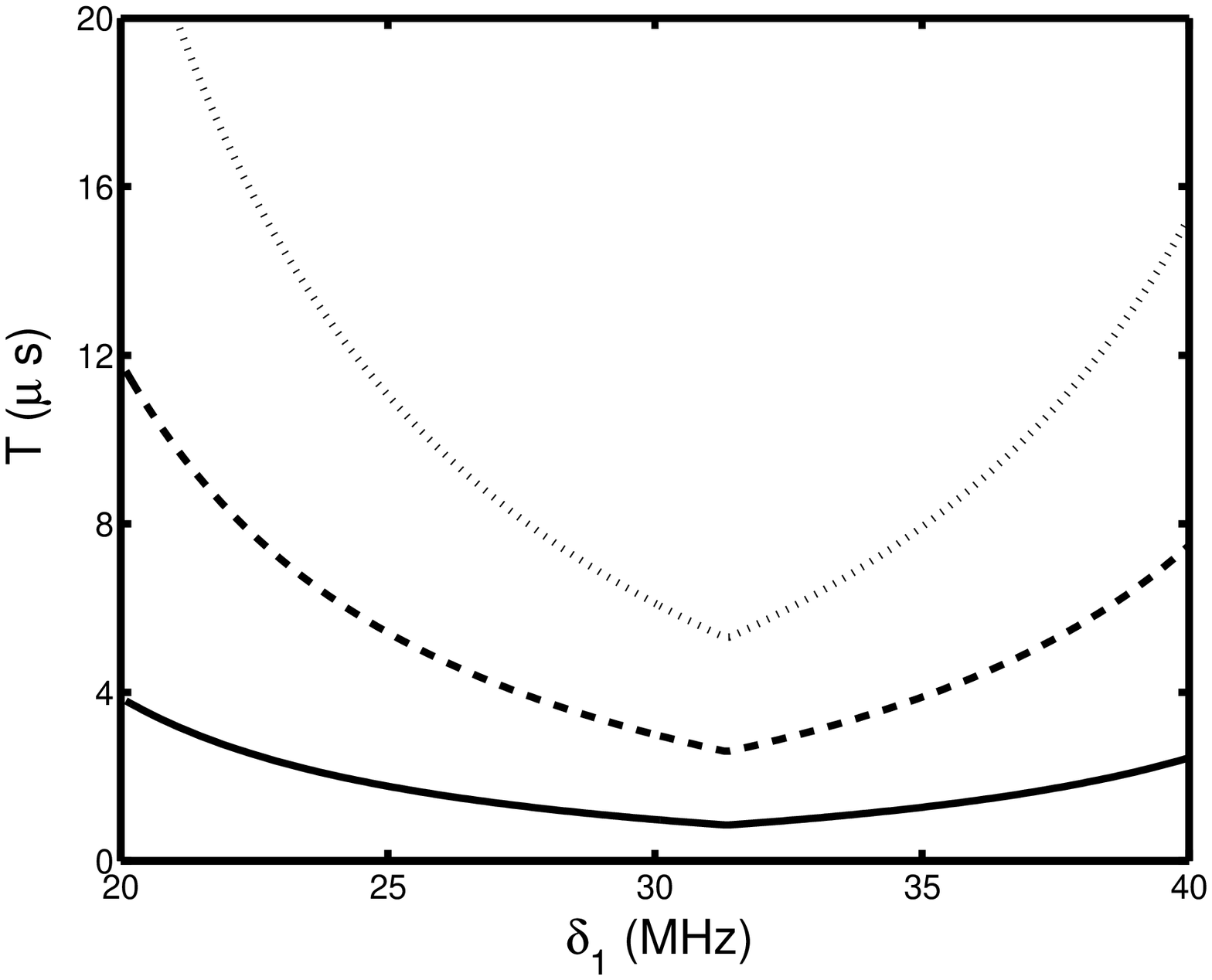}}
\end{picture}
\end{center}
\caption{ The implementation time $T=2\pi/\tilde{\Omega}$ with respect to $\delta_{1}$ in the implementation of our proposed 
CPHASE gate, based on Eq. (7). The dotted, dashed and solid curves correspond to gate fidelity $98\%$ (i.e., 
$\Omega_{A}/2=\Omega_{B}/2=\delta_{min}/10$), $95\%$ (i.e., $\Omega_{A}/2=\Omega_{B}/2=\delta_{min}/7$), and
$92\%$ (i.e., $\Omega_{A}/2=\Omega_{B}/2=\delta_{min}/5$), respectively. }
\label{Fig3}
\end{figure}
 

\begin{thebibliography}{22}
\bibitem{exp} Here we only briefly describe the transfer of spin message between two well polarized states in the readout stage of 
quantum information processing. The detailed  discussion for our readout scheme based on a large spin nanomolecular magnet 
$Fe_{8}$ or $Mn_{12}$ will be given in \cite {ft}.
\end{thebibliography}
\end{document}